\documentclass[aps,twocolumn,prl]{revtex4}
\usepackage{amsfonts}
\usepackage[dvips]{graphicx}
\usepackage[T1]{fontenc}
\usepackage{amsmath,amsbsy,amssymb,graphics}

\let\mathbf=\boldsymbol
\def\beginABC{\begin{subequations}}
\def\endABC{\end{subequations}}

\begin{document}

\title{Supersymmetry and Correlated Electrons in Graphene Quantum Hall Effect%
}
\author{Motohiko Ezawa}
\affiliation{Department of Physics, University of Tokyo, Hongo 7-3-1, 113-0033, Japan }
\date{\today}

\begin{abstract}
We present a supersymmetric description of the quantum Hall effect (QHE) in
graphene. The noninteracting system is supersymmetric separately at the
so-called K and K' points of the Brillouin zone corners. Its essential
consequence is that the energy levels and the Landau levels are different
objects in graphene QHE. Each energy level has a four-fold degeneracy within
the noninteracting theory. With the Coulomb interaction included, an
excitonic gap opens in the zero-energy state, while each nonzero energy
level splits into two levels since up-spin and down-spin electrons come from
different Landau levels. We argue the emergence of the plateaux at $\nu =\pm
(4n-2)$ for small magnetic field $B$ and at $\nu =0$, $\pm 1$, $\pm 2n$ for
large $B$ with $n$ natural numbers.
\end{abstract}

\maketitle

\textit{Introduction:} The quantum Hall effect (QHE) is one of the most
remarkable phenomena discovered in the last century\cite%
{BookDasSarma,BookEzawa}. Electrons, undergoing cyclotron motion in magnetic
field $B$, fill Landau levels successively. Each filled energy level
contributes one conductance quantum $e^{2}/\hbar $ to the Hall conductivity $%
\sigma _{xy}$. The Hall plateau develops at $\sigma _{xy}=\nu (e^{2}/h)$,
where $\nu $ is the filling factor. It tells us how many energy levels are
filled up. Hall plateaux have been observed at $\nu =1,2,3,\ldots $ in the
conventional semiconductor QHE.

\begin{figure}[h]
\includegraphics[width=0.48\textwidth]{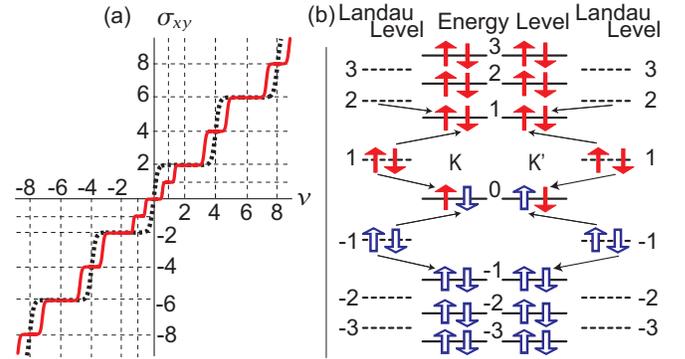}
\caption{{}(a) The QH conductivity in graphene. The dotted black curve shows
the sequence $\protect\nu =\pm 2,\pm 6,\pm 10,\cdots $, while the solid red
curve the sequence $\protect\nu =0,\pm 1,\pm 2,\pm 4,\cdots $. (b) The
energy level and the Landau level. The spin is indicated by a solid red
(open blue) arrow for electron and hole at the K and K' points. This energy
spectrum is a manifestation of SUSY. The $N$th energy level contains
up(down)-spin electrons from the $(N+1)$th Landau level and down(up)-spin
electrons from the $N$th Landau level at the K (K') point.}
\label{FigGraphLevel}
\end{figure}

Recent experimental developments have revealed unconventional QHE in graphene%
\cite{Nov1,Nov2,Zhang,Nov3}. The filling factors\cite{Ando,Gusynin,Peres}
form a series, $\nu =\pm 2,\pm 6,\pm 10,\cdots $, where the basic height in
the Hall conductance step is $4e^{2}/h$ [Fig.\ref{FigGraphLevel}(a)]. A
recent experiment\cite{Zhang06L} has shown a fine structure at $\nu =0,\pm
1,\pm 4$ when larger magnetic field is applied. In this Letter, we present a
mechanism how these unconventional filling factors arise on the basis of the
supersymmetric (SUSY) quantum mechanics\cite{Witten}. We also elucidate how
the graphene QHE is different from the semiconductor QHE.

The low-energy band structure of graphene is described by cones located at
two inequivalent Brillouin zone corners called the K and K' points. In these
cones, the two-dimensional energy dispersion relation is linear and the
dynamics can be treated as `relativistic' Dirac electrons\cite%
{Semenoff,Ajiki}, in which the Fermi velocity $v_{\text{F}}$ of the graphene
is substituted for the speed of light. The graphene system has the
pseudospin degree of freedom, where the electron at the K (K') point carries
the up (down) pseudospin.

The graphene QH system possesses a unique character that it has a
supersymmetry within the noninteracting theory\cite{EzawaM06B}. This follows
from the basic fact that the intrinsic Zeeman energy is precisely one half
of the cyclotron energy for Dirac electrons. It has two important
consequences [Fig.\ref{FigGraphLevel}(b)]; the emergence of the zero energy
state, and the degeneracy of the up-spin and down-spin states for each
nonzero energy level. Since this holds separately at the K and K' points,
each energy level has a four-fold degeneracy [Fig.\ref{FigGraphLevel}(b)],
and the noninteracting system has the SU(4) symmetry. The resulting series
is $\nu =\pm 2,\pm 6,\pm 10,\cdots $. In this paper we focus on the Coulomb
interactions, and explore how this degeneracy is modified. The zero energy
state is distinctive, since it contains both electrons and holes.
Electron-hole pairs form an excitonic condensation, producing an excitonic
gap. Hence, all states become gapful. According to the SUSY spectrum, a
single energy level contains up-spin and down-spin electrons belonging to
different Landau levels [Fig.\ref{FigGraphLevel}(b)]. It implies that their
wave functions and hence their Coulomb energies are different. The Coulomb
Hamiltonian, projected to a single energy level, possesses only the U(1)$%
\otimes $U(1)$\otimes $Z$_{2}$ symmetry. Thus, the Coulomb interaction leads
to the resolution of the four-fold degeneracy into two two-fold
degeneracies, each of which has the U(1) symmetry. As a result we obtain the
new series $\nu =0,\pm 1,\pm 4,\pm 8,\cdots $. Since all gap energies are
proportional to the Coulomb energy, or $\sqrt{B}$, this series is expected
to appear for large magnetic field.

\textit{SUSY spectrum:} We\ start with a concise review of the SUSY
description of graphene\cite{EzawaM06B}. Corresponding to the K and K'
points ($\tau =\pm $), we have two Dirac Hamiltonians%
\begin{equation}
H_{\text{D}}^{\tau }=v_{\text{F}}(\alpha _{x}(P_{x}-\tau \hbar K_{x})+\tau
\alpha _{y}(P_{y}-\tau \hbar K_{y}),  \label{DiracHamilD}
\end{equation}%
where $P_{i}\equiv -i\hbar \partial _{i}+eA_{i}$ is the covariant momentum,
and%
\begin{equation}
\alpha _{i}=\left( 
\begin{array}{cc}
0 & \sigma _{i} \\ 
\sigma _{i} & 0%
\end{array}%
\right)
\end{equation}%
with $\sigma ^{i}$ the Pauli matrix for spins and $\mathbf{K}=(4\pi /\sqrt{3}%
a,0)$. Here, $\pm \mathbf{K}$ represent the two Brillouin zone corners with $%
a$ the lattice constant. We assume a homogeneous magnetic field $\mathbf{B}=%
\mathbf{\nabla }\times \mathbf{A}=(0,0,-B)$ with $B>0$.

The $\mathbf{K}$ dependence is removed from the Hamiltonian (\ref%
{DiracHamilD}) by introducing the wave function $\varphi _{\tau }(\mathbf{x}%
)=e^{i\tau \mathbf{Kx}}\tilde{\varphi}_{\tau }(\mathbf{x})$. Then the
Hamiltonian (\ref{DiracHamilD}) is expressed as 
\begin{equation}
H_{\text{D}}^{\pm }=\left( 
\begin{array}{cc}
0 & Q_{\pm } \\ 
Q_{\pm } & 0%
\end{array}%
\right) ,  \label{DiracHamilB}
\end{equation}%
with $Q_{\pm }=v_{\text{F}}\left( \sigma _{x}P_{x}\pm \sigma
_{y}P_{y}\right) $. It is diagonalized\cite{Thaller92},%
\begin{equation}
H_{\text{D}}^{\pm }=\text{diag.}\left( \sqrt{Q_{\pm }Q_{\pm }},-\sqrt{Q_{\pm
}Q_{\pm }}\right) ,  \label{DiracHamilC}
\end{equation}%
where the negative component describes holes.

We consider the quantity 
\begin{equation}
H_{\text{P}}^{\pm }=Q_{\pm }Q_{\pm }=v_{\text{F}}^{2}\left( -i\hbar \nabla +e%
\mathbf{A}\right) ^{2}\mp e\hbar v_{\text{F}}^{2}\sigma _{z}B,
\label{PauliHamil}
\end{equation}%
where the direction of the magnetic field is effectively opposite at the K
and K' points. Since this has the same form as the Pauli Hamiltonian with
the mass $m^{\ast }=1/4v_{\text{F}}^{2}$ except for the dimension, we call
it the Pauli Hamiltonian for brevity. The salient feature of the \textit{%
relativistic} Dirac Hamiltonian is that its spectrum is mapped from that of
the \textit{nonrelativistic} Pauli Hamiltonian. Thus, the energy spectrum $%
\mathcal{E}_{n}$ of the Dirac Hamiltonian is constructed once we know the
one $E_{n}$ of the Pauli Hamiltonian.

In the Pauli Hamiltonian, the first term is the kinetic term while the
second term is the Zeeman term. It is fixed uniquely as an intrinsic
property of the Dirac theory: We may call it the intrinsic Zeeman effect.
The Landau level is created by electrons making cyclotron motion. In the
conventional QHE, since the Zeeman energy can be considered much smaller
than the Landau-level separation, we may treat it as a perturbation.
However, this is not the case in graphene.

The Hamiltonian (\ref{PauliHamil}) is a simplest example of the SUSY quantum
mechanics\cite{Witten}, where the superalgebra reads 
\begin{equation}
H_{\text{P}}^{\tau }=\frac{1}{2}\{Q_{\tau },Q_{\tau }\}
\end{equation}
and 
\begin{equation}
\left[ H^{\tau },Q_{\tau }\right] =0,
\end{equation}%
with $Q_{\tau }$ the supercharge ($\tau =\pm $). The energy eigenvalues of
the Dirac Hamiltonian $H_{\text{D}}^{\pm }$ are found\cite{EzawaM06B} to be 
\begin{equation}
\mathcal{E}_{0}^{+\uparrow }=\mathcal{E}_{0}^{-\downarrow }=0
\end{equation}
and 
\begin{equation}
\mathcal{E}_{n+1}^{+\uparrow }=\mathcal{E}_{n}^{+\downarrow }=\mathcal{E}%
_{n+1}^{-\downarrow }=\mathcal{E}_{n}^{-\uparrow }=\pm \hbar \omega _{c}%
\sqrt{n+1}
\end{equation}
for $n\geq 0$. There exists one zero-energy state only for up-spin
(down-spin) electrons at the K (K') point [Fig.\ref{FigGraphLevel}(b)]. The
existence of the zero energy state is an intriguing property of the SUSY\
theory, where the bosonic and fermionic zero-point energies are canceled out%
\cite{Witten}. The physical reason is that the intrinsic Zeeman splitting is
exactly as large as the Landau level separation in the Pauli Hamiltonian (%
\ref{PauliHamil}). The SUSY spectrum tells us that it is necessary to make a
clear distinction between the energy level and the Landau level. This point
has been overlooked in all previous literatures on graphene QHE.

The SUSY spectrum dictates that each energy level has a four-fold degeneracy
[Fig.\ref{FigGraphLevel}(b)]. The noninteracting theory has the SU(4)
symmetry. The resulting series is $\nu =\pm 2,\pm 6,\pm 10,\cdots $.

\textit{Projected Coulomb Hamiltonian:} We include the Coulomb interaction
to the noninteracting theory, and analyze how it affects the four-fold
degenerated electron states in the $N$th energy level. We project the
Coulomb interaction to the $N$th energy level, making a simple
generalization of the lowest-Landau-level projection\cite{Girvin84B}
familiar in the conventional QHE.

For definiteness we explicitly analyze the $N$th energy level with $N>0$.
The result for $N<0$ is obtained by the electron-hole symmetry without any
calculation. The case $N=0$ is analyzed in a similar way: See the discussion
on the exicitonic condensation we give soon after.

According to the SUSY spectrum [Fig.\ref{FigGraphLevel}(b)], the $N$th
energy level contains both up-spin (down-spin) electrons from the $(N$+$1)$%
th Landau level and down-spin (up-spin) electrons from the $N$th Landau
level at the K (K') point. In each energy level there exist four types of
electrons described by four different field operators $\psi _{N\tau
}^{\sigma }(\mathbf{x})$. They are expanded as 
\begin{align}
\psi _{N+}^{\uparrow }(\mathbf{x})& =\sum_{n}\varphi _{n+}^{N+1}(\mathbf{x}%
)c_{+}^{\uparrow }(n),  \notag \\
\psi _{N+}^{\downarrow }(\mathbf{x})& =\sum_{n}\varphi _{n+}^{N}(\mathbf{x}%
)c_{+}^{\downarrow }(n),  \notag \\
\psi _{N-}^{\uparrow }(\mathbf{x})& =\sum_{n}\varphi _{n-}^{N}(\mathbf{x}%
)c_{-}^{\uparrow }(n),  \notag \\
\psi _{N-}^{\downarrow }(\mathbf{x})& =\sum_{n}\varphi _{n-}^{N+1}(\mathbf{x}%
)c_{-}^{\downarrow }(n)
\end{align}%
in terms of the wave function 
\begin{equation}
\varphi _{n\tau }^{N}(\mathbf{x})=e^{i\tau \mathbf{Kx}}\langle \mathbf{x}%
|N,n\rangle ,
\end{equation}%
and the annihilation operator $c_{\tau }^{\sigma }(n)$ acting on the Fock
state $|N,n\rangle \equiv |N\rangle \otimes |n\rangle $ in the $N$th Landau
level with $n$ the Landau-site index.

We decompose the electron coordinate $\mathbf{x}=(x,y)$ into the guiding
center $\mathbf{X}=(X,Y)$ and the relative coordinate $\mathbf{R}%
=(R_{x},R_{y})$, $\mathbf{x}=\mathbf{X}+\mathbf{R}$, where $R_{x}=-P_{y}/eB$
and $R_{y}=P_{x}/eB$ with $\mathbf{P}=(P_{x},P_{y})$ the covariant momentum.
The projection is to quench the motion in the relative coordinate.

In the $N$th energy level, by decomposing the guiding center $\mathbf{X}$
and the relative coordinate $\mathbf{R}$, the density operator reads%
\begin{equation}
\rho _{N}(\mathbf{q})=\sum_{\sigma \tau \tau ^{\prime }}\psi _{N\tau
}^{\sigma \dag }(\mathbf{q})\psi _{N\tau ^{\prime }}^{\sigma }(\mathbf{q}%
)=\sum_{\sigma \tau \tau ^{\prime }}F_{\tau \tau ^{\prime }}^{\sigma }(%
\mathbf{q})\hat{D}_{\tau \tau ^{\prime }}^{\sigma \sigma }(\mathbf{q}),
\label{G-Densi}
\end{equation}%
where $\hat{D}_{\tau \tau ^{\prime }}^{\sigma \sigma ^{\prime }}(\mathbf{q})$
is the projected density\cite{Ezawa05D},%
\begin{equation}
\hat{D}_{\tau \tau ^{\prime }}^{\sigma \sigma ^{\prime }}(\mathbf{q})=\frac{1%
}{2\pi }\sum_{mn}\langle m|e^{-i[\mathbf{q}+\tau \mathbf{K}-\tau ^{\prime }%
\mathbf{K}]\mathbf{X}}|n\rangle c_{\tau }^{\sigma \dagger }(m)c_{\tau
^{\prime }}^{\sigma ^{\prime }}(n),  \label{G-ProjeDensi}
\end{equation}%
and $F_{\tau \tau ^{\prime }}^{\sigma }(\mathbf{q})$ is the form factor%
\begin{align}
F_{++}^{\downarrow }\left( \mathbf{q}\right) & =F_{--}^{\uparrow }\left( 
\mathbf{q}\right) =\langle N|e^{-i\mathbf{qR}}|N\rangle ,  \notag \\
F_{++}^{\uparrow }\left( \mathbf{q}\right) & =F_{--}^{\downarrow }\left( 
\mathbf{q}\right) =\langle N+1|e^{-i\mathbf{qR}}|N+1\rangle ,  \notag \\
F_{+-}^{\uparrow }\left( \mathbf{q}\right) & =F_{-+}^{\downarrow }\left( 
\mathbf{q}\right) =\langle N+1|e^{-i(\mathbf{q-K)R}}|N\rangle ,  \notag \\
F_{-+}^{\uparrow }\left( \mathbf{q}\right) & =F_{+-}^{\downarrow }\left( 
\mathbf{q}\right) =\langle N|e^{-i(\mathbf{q}+\mathbf{K)R}}|N+1\rangle .
\label{FormFactoEL}
\end{align}%
The set of these form factors is a direct consequence of the SUSY spectrum.
For instance, $F_{+-}^{\uparrow }$ represents the transfer of the up-spin
electron ($\sigma =\uparrow $) from the K' point ($\tau =-$) to the K point (%
$\tau =+$). In this process an electron in the $N$th Landau level is moved
to the $(N$+$1)$th Landau level. Thus, the form factor necessarily mixes the 
$N$th Landau level and the $(N$+$1)$th Landau level. This is an essential
difference from the conventional QHE, where only one Landau level is
involved. The form factors are explicitly given by using\cite{Ando74JPSJ}%
\begin{align}
\langle N\!+\!M|e^{i\mathbf{qR}}|N\rangle & =\frac{\sqrt{N!}}{\sqrt{(N+M)!}}%
\left( \frac{\ell _{B}q}{\sqrt{2}}\right) ^{M}L_{N}^{M}\left( \frac{\ell
_{B}^{2}\mathbf{q}^{2}}{2}\right)  \notag \\
& \hspace{2.9cm}\times e^{-\frac{1}{4}\ell _{B}^{2}\mathbf{q}^{2}}
\label{FormFactoLague}
\end{align}%
for $M\geq 0$ in terms of the associated Laguerre polynomial.

The projected Coulomb Hamiltonian is%
\begin{equation}
H_{N}=\pi \sum_{\tau \tau ^{\prime }\sigma }\sum_{\lambda \lambda ^{\prime
}\sigma ^{\prime }}\int {\!d^{2}q\,}V_{\tau \tau ^{\prime }\lambda \lambda
^{\prime }}^{\sigma \sigma ^{\prime }}(\mathbf{q})\hat{D}_{\tau \tau
^{\prime }}^{\sigma \sigma }(-\mathbf{q})\hat{D}_{\lambda \lambda ^{\prime
}}^{\sigma ^{\prime }\sigma ^{\prime }}(\mathbf{q}),  \label{ProjeHamilN}
\end{equation}%
where 
\begin{equation}
V_{\tau \tau ^{\prime }\lambda \lambda ^{\prime }}^{\sigma \sigma ^{\prime
}}(\mathbf{q})=V(\mathbf{q})F_{\tau \tau ^{\prime }}^{\sigma }(-\mathbf{q}%
)F_{\lambda \lambda ^{\prime }}^{\sigma ^{\prime }}(\mathbf{q})
\end{equation}
with the Coulomb potential $V(\mathbf{q})=e^{2}/4\pi \varepsilon |\mathbf{q}%
| $.

The Coulomb Hamiltonian $H_{N}$ is invariant under two U(1) transformations,%
\begin{equation}
\left( 
\begin{array}{c}
c_{+}^{\uparrow } \\ 
c_{-}^{\uparrow }%
\end{array}%
\right) \rightarrow e^{i\alpha }\left( 
\begin{array}{c}
c_{+}^{\uparrow } \\ 
c_{-}^{\uparrow }%
\end{array}%
\right) ,\quad \left( 
\begin{array}{c}
c_{+}^{\downarrow } \\ 
c_{-}^{\downarrow }%
\end{array}%
\right) \rightarrow e^{i\beta }\left( 
\begin{array}{c}
c_{+}^{\downarrow } \\ 
c_{-}^{\downarrow }%
\end{array}%
\right) ,  \label{G-U1}
\end{equation}%
with two arbitrary constants $\alpha $ and $\beta $. Additionally it is
invariant under the Z$_{2}$ transformation,%
\begin{equation}
\left( 
\begin{array}{c}
c_{+}^{\uparrow } \\ 
c_{-}^{\uparrow }%
\end{array}%
\right) \longleftrightarrow \left( 
\begin{array}{c}
c_{-}^{\downarrow } \\ 
c_{+}^{\downarrow }%
\end{array}%
\right) ,  \label{G-Z2}
\end{equation}%
as corresponds to the fact that the magnetic field is opposite at the K and
K' points. The symmetry is quite small because of the Landau-level mixing in
a single energy level, as follows from the SUSY spectrum of the
noninteracting theory.

\begin{figure}[h]
\includegraphics[width=0.48\textwidth]{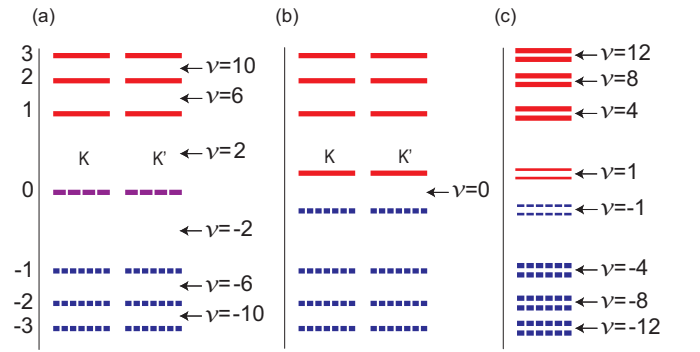}
\caption{{}A schematic illustration of the energy spectrum. (a) Coulomb
interactions are neglected. All states are four-fold degenerated. (b)
Excitonic condensation is taken into account at the zero-energy level. The
four-fold degeneracy splits into two two-fold degeneracies with the gap
energy (\protect\ref{GapAtZero}). Electron (solid red line) and hole (dotted
blue line) states are separated. (c) All Coulomb interactions are taken into
account. Each level splits into two subbands. The U(1) symmetry is exact in
each two-fold degenerated level.}
\label{FigSplit}
\end{figure}

\textit{Excitonic condensation:} It is necessary to treat the zeroth energy
level ($N=0$) separately from the others, since it contains both electrons
and holes. They come from the $N$th Landau levels with $N=\pm 1$. Due to the
Coulomb attraction, electron-hole pairs are expected to make bound states
and condense into the spin-singlet excitonic states.

It is sufficient to investigate only electron-hole pairs at the K and K'
points separately, because an exciton composed of an electron and a hole
belonging to different Dirac cones is fragile. At the K point their field
operators are%
\begin{equation}
\psi _{\text{e}}(\mathbf{x})=\sum_{n}\varphi _{n+}^{1}(\mathbf{x})c_{+}^{%
\text{e}\uparrow }(n),\quad \psi _{\text{h}}(\mathbf{x})=\sum_{n}\varphi
_{n+}^{-1}(\mathbf{x})c_{+}^{\text{h}\downarrow }(n).
\end{equation}%
The dominant electron-hole interaction is%
\begin{equation}
H_{N}=\pi \int {\!d^{2}q\,}V(\mathbf{q})\psi _{\text{e}}^{\dag }(-\mathbf{q}%
)\psi _{\text{e}}(-\mathbf{q})\psi _{\text{h}}^{\dag }(\mathbf{q})\psi _{%
\text{h}}(\mathbf{q}).
\end{equation}%
The analysis of this Hamiltonian is similarly done as in the BCS theory.
However, the kinetic term is absent since it is quenched in each Landau
level. This simplifies the analysis considerably. In the mean-field
approximation, the singlet excitonic gap function satisfies the gap equation,%
\begin{equation}
\Delta \left( \mathbf{k}\right) =\int d^{2}k^{\prime }V(\mathbf{k}^{\prime }-%
\mathbf{k)}e^{-l_{B}^{2}(\mathbf{k}^{\prime }-\mathbf{k})^{2}/2}\tanh \frac{%
\Delta (\mathbf{k}^{\prime })}{2k_{\text{B}}T},
\end{equation}%
where $k_{\text{B}}$ is the Boltzmann factor, and the suppression factor $%
e^{-l_{B}^{2}(\mathbf{k}^{\prime }-\mathbf{k})^{2}/2}$ has arisen from the
projection of the Coulomb interaction. In the limit $T\rightarrow 0$, the
zero-momentum gap $\Delta \left( 0\right) $ is given by%
\begin{equation}
\left. \Delta \left( 0\right) \right\vert _{T=0}=\pi \sqrt{2\pi }\left(
e^{2}/4\pi \varepsilon l_{B}\right) .  \label{GapAtZero}
\end{equation}%
The excitonic condensation resolves the electron-hole degeneracy in the
zero-energy state. As a result the four-fold degenerated levels split into
two two-fold degenerated levels [Fig.\ref{FigSplit}(b)] with the gap energy (%
\ref{GapAtZero}). This leads to a new plateau at $\nu =0$.

\textit{Pseudospin asymmetry:} We can treat electrons and holes separately
since a gap has opened between the electron and hole bands [Fig.\ref%
{FigSplit}(b)]. Each energy level is four-fold degenerated within the
noninteracting model with the SU(4) symmetry. However, the Coulomb
interaction breaks it explicitly into U(1)$\otimes $U(1)$\otimes $Z$_{2}$.
We now argue that plateaux emerges at $\nu =\pm 1,\pm 2n$ by this symmetry
reduction [Fig.\ref{FigSplit}(c)].

We take the Hartree-Fock trial function for the $N$th energy level ($N\neq 0$%
) by requiring the symmetry,\beginABC%
\begin{align}
|\Phi _{N}^{\uparrow }\rangle & =\prod\limits_{n}\left( u^{\uparrow
}c_{+}^{\uparrow \dagger }\left( n\right) +v^{\uparrow }c_{-}^{\uparrow
\dagger }\left( n\right) \right) \left\vert 0\right\rangle , \\
|\Phi _{N}^{\downarrow }\rangle & =\prod\limits_{n}\left( u^{\downarrow
}c_{+}^{\downarrow \dagger }\left( n\right) +v^{\downarrow
}c_{-}^{\downarrow \dagger }\left( n\right) \right) \left\vert
0\right\rangle ,
\end{align}%
\endABC
with $|u^{\sigma }|^{2}+|v^{\sigma }|^{2}=1$. They transforms properly under
the U(1) transformation (\ref{G-U1}), and exchange themselves under the Z$%
_{2}$ transformation (\ref{G-Z2}).

It is easy to determine $u^{\sigma }$ and $v^{\sigma }$ by minimizing the
Coulomb energy $\langle \Phi _{N}^{\sigma }|H_{\text{C}}|\Phi _{N}^{\sigma
}\rangle $. The result says%
\begin{equation}
u^{\uparrow }=v^{\downarrow }=e^{i\theta }\sin \alpha ,\quad v^{\uparrow
}=u^{\downarrow }=e^{-i\theta }\cos \alpha ,
\end{equation}%
where $\alpha $ is a certain constant given in terms of integrals over
various form factors together with the Coulomb potential. The arbitrary
phase $\theta $ assures the U(1) symmetry. It follows that $|u^{\uparrow
}|>|v^{\uparrow }|$, leading to the pseudospin asymmetry, because the
Coulomb energy of an electron in higher Landau level is lower. These two
states are degenerate, 
\begin{equation}
\langle \Phi _{N}^{\uparrow }|H_{\text{C}}|\Phi _{N}^{\uparrow }\rangle
=\langle \Phi _{N}^{\downarrow }|H_{\text{C}}|\Phi _{N}^{\downarrow }\rangle
,
\end{equation}
due to the Z$_{2}$ invariance (\ref{G-Z2}).

To study the zeroth energy level, we take the trial function 
\begin{equation}
|\Phi _{0}\rangle =\prod_{n}\left( uc_{+}^{\uparrow \dagger }\left( n\right)
+vc_{-}^{\downarrow \dagger }\left( n\right) \right) \left\vert
0\right\rangle
\end{equation}
with $|u|^{2}+|v|^{2}=1$. By minimizing the energy we find either $u=0$ or $%
v=0$, reflecting the Z$_{2}$ symmetry.

We have so far taken the intrinsic Zeeman effect into account. However,
there may be an additional Zeeman effect in graphene, which make these two
states split explicitly. Even without such an extrinsic Zeeman effect,
driven by the Coulomb exchange interaction, the spontaneous breakdown of the
Z$_{2}$ symmetry turns the system into a QH ferromagnet\cite{Moon95B}. In
any case the excitation gap is of the order of the typical Coulomb energy.
This explains the emergence of plateaux at $\nu =\pm 1,\pm 2,\pm 4,\cdots $.

\textit{Discussions:} We have presented a SUSY description of the QHE in
graphene. It has an important consequence that the energy levels and the
Landau levels are quite different objects in graphene [Fig.\ref%
{FigGraphLevel}(b)]. One energy level contains electrons coming from two
neighboring Landau levels. This is evidenced in the Coulomb energy through
the form factor (\ref{FormFactoEL}). We have derived the Coulomb Hamiltonian
(\ref{ProjeHamilN}), as dictated by the SUSY\ spectrum.

Our formalism is considerably different from those assumed by previous
authors. Let us comment on them. First of all, Alicia and Fisher\cite{Fisher}
have not taken into account the projection of the Coulomb interaction. They
have also ignored the Landau-level mixing in a single energy level. Nomura
and MacDonald\cite{Nomura} made only an explicit analysis of the SU(4)
symmetric term, which is extracted from (\ref{G-Densi}) as%
\begin{equation}
\rho _{N}^{\text{SU(4)}}(\mathbf{q})=\frac{1}{2}\{F_{N}(\mathbf{q})+F_{N+1}(%
\mathbf{q})\}\sum_{\sigma \tau }\hat{D}_{\tau \tau }^{\sigma \sigma }\left( 
\mathbf{q}\right) ,
\end{equation}%
where we have set $F_{N}(\mathbf{q})=\langle N|e^{-i\mathbf{qR}}|N\rangle $.
We should note that the SU(4) noninvariant term is as large as the invariant
term. This is essentially different from the conventional QH system, where
the noninvariant term can be made arbitrarily small by controlling the
external parameters such as the layer separation $d$, the tunneling gap $%
\Delta _{\text{SAS}}$ and the magnetic $g$-factor. Finally, the spin degree
of freedom has been ignored by Goerbig et al.\cite{Goerbig}, where $%
F_{N}\left( \mathbf{q}\right) $ and $F_{N+1}\left( \mathbf{q}\right) $
appear only in the SU(4) symmetric combination in the density operator.

We have started with the SU(4) symmetry in the noninteracting theory, where
gaps open at $\nu =\pm 2,\pm 6,\pm 10,\cdots $. The Coulomb interaction
breaks it explicitly into the U(1)$\otimes $U(1)$\otimes $Z$_{2}$ symmetry.
As a result, new gaps open at $\nu =\pm 1,\pm 4,\pm 8,\cdots $. The
remaining problem is whether a further resolution of the degeneracy may
occur.

Here, we recapture a similar problem in the spin-frozen bilayer QH system%
\cite{BookEzawa}. Let us ignore the tunneling gap $\Delta _{\text{SAS}}$.
Then, there is the pseudospin SU(2) symmetry in the noninteracting theory,
which is broken to the U(1) symmetry explicitly by the capacitance effect.
Here, the exchange Coulomb interaction generates the pseudospin wave, which
is the Goldstone mode associated with spontaneous breakdown (SSB) of a
continuous symmetry, and turns the bilayer system into the pseudospin QH
ferromagnet\cite{Moon95B}. Based on this analogy Nomura and MacDonald\cite%
{Nomura} has concluded a spontaneous development of the SU(4) QH ferromagnet
in graphene.

We question why the SSB of a continuous symmetry is possible at finite
temperature in spite of the Mermin-Wagner theorem in the conventional QHE.
This is because there exists the tunneling gap $\Delta _{\text{SAS}}\neq 0$
in the actual bilayer system, which gives a gap to the Goldstone mode. This
is not the case in the graphene QHE. Graphene is an ideal two-dimensional
system, and furthermore the U(1) symmetry is exact. There is no external
parameter which breaks it explicitly to make the Mermin-Wagner theorem
inapplicable. We conclude that the SSB of the U(1) symmetry cannot occur in
graphene, so that the plateaux at $\nu =\pm 3,\pm 5,\pm 7,\cdots $ will not
emerge.

The author is grateful to Professors T. Ando, Y. Iye and Z.F. Ezawa for
fruitful discussions on the subject.

\end{document}